\documentclass[conference]{IEEEtran}
\IEEEoverridecommandlockouts

\usepackage{cite}
\usepackage[ruled,vlined]{algorithm2e}
\usepackage{amsmath,amssymb,amsfonts}
\usepackage{algorithmic}
\usepackage{graphicx}
\usepackage{textcomp}
\usepackage{xcolor}

\usepackage{makecell}

\begin{document}

\title{Hardware Acceleration of LLMs: A comprehensive survey and comparison}

\author{\IEEEauthorblockN{Nikoletta Koilia}
\IEEEauthorblockA{\textit{Department of Electrical} \\
\textit {and Electronics Engineering} \\
\textit{University of West Attica}\\
Athens, Greece \\
eee19387106@uniwa.gr}
\and
\IEEEauthorblockN{Christoforos Kachris}
\IEEEauthorblockA{\textit{Department of Electrical} \\
\textit {and Electronics Engineering} \\
\textit{University of West Attica}\\
Athens, Greece \\
kachris@uniwa.gr}
}



\maketitle

\begin{abstract}

Large Language Models (LLMs) have emerged as powerful tools for natural language processing tasks, revolutionizing the field with their ability to understand and generate human-like text. In this paper, we present a comprehensive survey of the several research efforts that have been presented for the acceleration of transformer networks for Large Language Models using hardware accelerators.
 The survey presents the frameworks that have been proposed and then performs a qualitative and quantitative comparison regarding the technology, the processing platform (FPGA, ASIC, In-Memory, GPU), the speedup, the energy efficiency, the performance (GOPs), and the energy efficiency (GOPs/W) of each framework. 
The main challenge in comparison is that every proposed scheme is implemented on a different process technology making hard a fair comparison. The main contribution of this paper is that we extrapolate the results of the performance and the energy efficiency on the same technology to make a fair comparison; one theoretical and one more practical. We implement part of the LLMs on several FPGA chips to extrapolate the results to the same process technology and then we make a fair comparison of the performance.

\end{abstract}





\maketitle

\section{Introduction}

Modeling human language on a large scale is a complex process that has taken decades to develop. It started in 1950 with Claude Shannon, who applied information theory to human language. Since then, tasks like translation and speech recognition have advanced significantly.

Artificial Intelligence (AI) and Machine Learning (ML) are key to this progress. ML, a subset of AI, allows computers to learn from data. ML models are either supervised (making predictions) or unsupervised. This thesis focuses on supervised models, which predict and compare values to minimize error through optimization.

Deep Learning models are divided into Generative (creating new data) and Discriminative (distinguishing data types). Generative AI, a subset of deep learning, uses neural networks to process labeled and unlabeled data. Large Language Models (LLMs) help understand characters, words, and texts.

In 2017, transformers revolutionized language modeling. Transformers, a type of neural network, handle long-term text dependencies using an attention mechanism. Google created the first transformer model for text translation in 2017. Transformers have since evolved, improving attention mechanisms and architectures.

ChatGPT, a notable LLM, predicts text continuations and performs tasks like answering questions, summarizing texts, and more. It uses probability distributions to generate various text forms based on user requests.

\subsection{LLMs}
Large Language Models (LLMs) are extensive, general-purpose models that can be pre-trained and adapted for specific tasks. They solve common language problems such as text classification, question answering, summarization, and text generation in various domains.

LLMs are "general-purpose" because they handle diverse tasks and "large" due to their massive training datasets and numerous parameters. These models have multiple neural network layers with adjustable weights that learn to predict the next word in a sentence during training.

The number of parameters indicates the model's complexity and capacity. Weights, adjusted during training, connect neurons in different layers, influencing the model's performance.

Transformers, a type of LLM, consist of an encoder and a decoder. The encoder has six layers, each with Multi-Head Self-Attention and a feed-forward network. The decoder has six layers, including an additional multi-head attention layer over the encoder's output.

The attention mechanism maps queries and key-value pairs to outputs, with positional encoding adding information about character positions. This architecture enables transformers to handle long-term dependencies in text effectively

\subsection{Encoder-Decoder}
The encoder-decoder architecture is central to Large Language Models (LLMs) and designed to process and generate sequences. This architecture has two stages:

\begin{itemize}
\item \textbf{Encoder:} The input (e.g., natural language) is transformed into a vector representation that encapsulates the meaning of the input.

\item \textbf{Decoder:} The decoder takes this vector representation and generates an output sequence, such as a translation into another language.
\end{itemize}

\subsection{Attention Mechanism}
The Attention Mechanism is vital in modern machine learning, especially in transformers, improving sequence processing tasks like translation and text generation. It connects both the encoder and decoder stages.
The Attention Mechanism includes two further mechanisms: Multi-Head Attention and Self-Attention.
The former focuses attention on different parts of the input simultaneously, allowing the model to recognize complex patterns and relationships in the input data. The latter captures dependencies and relationships between tokens regardless of their distance. It uses three matrices: Query (Q), Key (K), and Value (V). These matrices determine how much attention each token should give to another, enhancing the quality of translations and other sequence-based tasks.

\subsection{Related work}
Until now there is not any comprehensive survey on the hardware accelerators to speed-up the most computational intensive tasks of Transformers. In \cite{survey_transformer}, a survey has presented a survey on the hardware acceleration of transformer networks for autonomous driving. The paper presents several efforts on the acceleration of tasks such as object detection, 3D segmentation, and lane detection. 

In 2022, Huang et al. presented a survey on hardware acceleration for transformers \cite{hw_survey_huang}. The paper was mostly focused on the the transformer model compression algorithm based on the hardware accelerator and was limited mostly on FPGA-based implementation. 

In 2023, Emani et al \cite{2023comprehensive} presented a comprehensive performance study of LLMs on several computing platforms and evaluated their performance characteristics for these models.

In this paper, we present a comprehensive survey of the several research efforts that have been presented for the acceleration of transformer networks for Large Language models and NLP using hardware accelerators. The survey presents the frameworks that have been proposed and then performs a qualitative and quantitative comparison regarding the technology, the processing platform (GPU, FPGA, ASIC, In-Memory), the performance, and the energy efficiency of each framework. First, we present the accelerators based on FPGAs, then we present the accelerators targeting GPUs and finally accelerators ported on ASICs and In-memory architectures. 

The main contributions of this papers are the followings:

\begin{itemize}
    \item An extensive survey of hardware acceleration of LLM using FPGA, ASICs, In-memory architectures and GPUs. 
    \item A comparison in terms of performance (GOPs), energy efficiency (GOPs/W) and speedup. 
    \item An extrapolation of the features to the same technology for a fair comparison in terms of performance and energy efficiency. 
   
\end{itemize}

\section{FPGA-based accelerators}


\subsection{FTRANS}

In 2020, Li et al \cite{2020_ftrans} presented a hardware acceleration framework, called FTRANS, that was targeting the acceleration of transformer-based large scale language representations. It focuses on compression and acceleration to address computing and storage requirements, achieving up to 16 times compression with minimal accuracy loss through a Block Circulant Matrix (BCM) based weight model. The model significantly improves speed and energy efficiency, surpassing CPU and GPU implementations, with a comparison showing FTRANS is 81x faster and 9x more energy-efficient than alternatives, specifically compared to the GPU processor RTX5000 using VCU118 (16nm). The accelerator achieves a performance rate of 170 GOPs and an energy efficiency rate of 6.8 GOPs/W.

\subsection{Multi-Head Attention}

In 2020, Lu et al. presented an FPGA based architecture for the acceleration of the most computationally intensive parts of transformer networks \cite{2020_multihead}. In their work they propose a novel hardware accelerator for two key components, i.e., the multi-head attention (MHA) ResBlock and the position-wise feed-forward network (FFN) ResBlock, which are the two most complex layers in the Transformer.

The proposed framework is implemented on a Xilinx FPGA. Based on the performance evaluation the proposed design achieves a speed-up of 14.6× compared to a V100 GPU.

\subsection{FPGA NPE}

In 2021, Khan et al. presented an FPGA acceleration for language models called NPE. \cite{2021_FPGA_NPE}. The NPE architecture consists of an instruction control unit (ICU), a memory read unit (MRU), a memory write unit (MWU), a matrix multiply unit (MMU), and a nonlinear vector unit (NVU).

NPE was implemented on Xilinx Zynq Z-7100 FPGA board clocked at 200 MHz. NPE is compared with other frameworks like FTRANS and implementation on CPU and GPU. Although that there is not any significant speedup compared to other computing platforms, the main advantage is the energy efficiency. NPE achieves around 4× better energy efficiency over CPU (i7-8700k) and 6× over GPU (RTX 5000). 

\subsection{Column Balanced Block Pruning}

In 2021, Peng et al. presented a novel scheme on accelerating Transformer networks using column balanced block-wise pruning \cite{2021_pruning}. The column balanced block-wise pruning combines the key features of both bank balanced pruning and block-wise pruning. The column balanced block-wise pruning ranks the blocks’ L2 norm by each column to get the pruning thresholds and prunes blocks for each column.

The proposed framework has been implemented on different hardware platforms (Intel i5-5257U (2.7 GHZ) CPU, Nvidia Jetson TX2 GPU, and Xilinx Alveo U200 FPGA) for further comparison of latency and throughput. The experimental results showed that the FPGA platform achieves a 11× speed up compared to the CPU platform and 2× speed up compared to the GPU platform.

\subsection{Compressed Block Row}

In 2021, Panjie Qi et al, presented an acceleration framework that combines balanced model compression at the algorithm level with an FPGA implementation optimization at the hardware level \cite{column_balanced}. In their work, they propose an effective sparse matrix storage structure for block-balanced pruning, known as Compressed Block Row (CBR), and their hardware design includes an accelerator for sparse models. Moreover, they present a performance analytic methodology for evaluating accelerator performance. The experiments demonstrate that their CBR format outperforms conventional formats and saves substantial storage space. 

The proposed framework is implemented on a Xilinx ALveo U200 FPGA. Based on the performance evaluation the proposed design achieves a speed-up of 38x compared to a Nvidia Guardo RTX 6000.

\subsection{ViA}

In 2022, Teng Wang et al, presented ViA \cite{2021_via}, an FPGA-based accelerator architecture for Vision Transformers (ViT), featuring a memory recognition unit, a memory write unit, and processing elements like the NSA self-attention module and MLP. It proposes data partitioning strategies to enhance efficiency and reduce dependency. ViA's FPGA implementation significantly outperforms CPUs, GPUs, and previous FPGA accelerators, achieving 60x the speed and 5x the energy efficiency of alternatives like the Nvidia Tesla V100 and Alveo U50 (16nm). ViA reaches an acceleration rate of 309.6 GOPs and an energy efficiency rate of 7.9 GOPs/W.

\subsection{FPGA DFX}

In 2022, Hong et al. presented DFX \cite{2022_DFX} for the acceleration of the transformer networks used in LLMs. Similarly to NPE, the DFX architecture proposed a modular architecture consisting for several computer core for the acceleration of the transformer networks.

For the evaluation, DFX has been implemented on an Intel Xeon Gold 6226R CPU with four Xilinx Alveo U280 data center acceleration cards. DFX achieves an average of 3.8x throughput and 4x higher energy efficiency compared to the GPU appliances.

\subsection{STA}

In 2022, Chao Fang et al, presented the Sparse Transformer Accelerator (STA) on FPGA to address the high computational demands of transformer models\cite{sta}. Utilizing an N
structure, the STA minimizes operations and memory size while enhancing performance. The design includes a unified matrix multiplication mechanism, a Softmax module, and a Dense Matrix Multiplication Engine (DMME), implemented on an Intel Arria 10 SX660 device. It significantly improves energy efficiency and reduces latency compared to previous FPGA methods.

The STA is divided into STA-4 and STA-8 subcategories. STA-4 achieves 6.7 times better performance and is 10 times more energy-efficient than other models, with an acceleration rate of 392.8 GOPs and energy efficiency of 33.6 GOPs/W, using Nvidia RTX 2080Ti for comparison. STA-8, while slightly less performant with 4.4x better performance, offers 12.3x better energy efficiency, achieving an acceleration rate of 523.8 GOPs and energy efficiency of 41.2 GOPs/W. 

\subsection{FPGA OPU}

In 2023, Bai et al. proposed another scheme for the acceleration of transformer networks called Overaly OPU \cite{2023_OPU}. They propose a configurable computation unit to support the inference of diverse networks. Specifically, they propose 48 processing elements (PEs) that are configured for the acceleration of the transformer networks. The output stage of the adder tree can be switched during the inference process. That way, data from forwarding modules can flow through the computation unit in a pre-defined connection state. The proposed scheme achieves 5x-15× speedup compared with a CPU, 1.1-2.9× speedup compared with GPU (RTX 3090), and, 1.10-2.5× speedup compared with the other FPGA accelerators such as NPE \cite{2021_FPGA_NPE}.

\subsection{FPGA acceleration of Transformer networks}


In 2022, Tzanos et al, presented a high-performance hardware accelerator for the transformer networks \cite{tzanos}.  Transformer networks use a technique called attention. The attention, adopted by the field of neuroscience, is the ability to be able to selectively concentrate on specific data while ignoring other data of the environment. In deep learning we imitate this technique through attention mechanisms and one way to achieve this is to encode a sequence not into a single fixed vector but to create a model that produces a vector for each output step by adding a set of weights which will later be optimized.

The performance evaluation showed that the proposed framework can achieve 2.3x system speedup for the BERT model compared to a 40-thread processor and 80.5x speed-up over a single-core CPU. 

\subsection{FlexRun}

In 2023, Hur at al. presented an FPGA-based accelerator to speedup the diverse and complex NLP models, called FlexRun \cite{hw_fpga_flexrun}. The paper is focused on accelerating both Recurrent Neural Networks (RNNs) models such as SRNN or long short term memory (LSTM) and attention-based NLP models, such as Transformer, and GPT2. 
\hbadness=99999

For evaluation, they compare FlexRun with Intel’s Brainwave-like architecture on a Stratix-10 GX FPGA and a Tesla V100 GPU with tensor cores enabled. Compared to the FPGA baseline, FlexRun achieves an average speedup of 1.59× on various configurations of BERT. For GPT2, FlexRun gets 1.31× average speedup. Next, when comparing to the GPU implementation, FlexRun improves the performance by 2.79× and 2.59× for BERT and GPT2, respectively. 

\subsection{HPTA}
In 2023, Yuntao Han and Qiang Liu presented the High-Performance Transformer Accelerator (HPTA) \cite{hpta}, leveraging a custom multiplication matrix, adder tree, and memory subsystem. It can handle various types of transformers used in Natural Language Processing (NLP) and Computer Vision (CV). The performance of HPTA was evaluated against CPU, GPU, and other FPGA implementations. The results showed significant improvements in speed and energy efficiency for both BERT and Swin Transformer models. Compared to CPU and GPU, HPTA processed BERT up to 44x faster and 175x more energy-efficiently. It was also 1.8x faster than previous FPGA accelerators

\subsection{Swin}
In 2023, Zhiyang Liu, Zhenhua Ren, and Pengyu Yin developed an accelerator for the Swin Transformer in computer vision tasks, addressing hardware acceleration challenges with large images \cite{swin}. The architecture includes computation units for GELU and Softmax, allowing Swin Transformer Block execution in one cycle and improving efficiency by replacing Layer Normalization (LN) with Batch Normalization (BN). It offers significant speed and energy efficiency improvements over CPU and GPU. The accelerator is categorized into Swin-T, Swin-S, and Swin-B. Swin-T is 1.8x faster and 20.5x more energy-efficient, Swin-S is 1.7x faster and 18.6x more energy-efficient, and Swin-B is 4.4x faster and 14.6x more energy-efficient compared to the Nvidia GeForce RTX 2080Ti. The acceleration rates are 431.2, 403.5, and 436.4 GOPs for Swin-T, Swin-B, and Swin-S, respectively.

\subsection{Zhongyo Zhao}
In 2023, Zhongyo Zhao presented an accelerator that uses an Output Block Storing (OBS) data handling method to efficiently execute transformer models for object recognition \cite{zhao_cao}. The proposed method involves dividing the inputs and allocating weights into small block matrices to reduce memory access for input data and weights. Additionally, the OBS data flow maintains usage rates by collecting partial sums, while slightly reducing them compared to the output block data flow. This results in improved overall energy efficiency. The accelerator implements this data flow and achieves a processing rate of 728.3 GOPs and an energy efficiency of 58.31 GOPs/W, surpassing previous CNN-based accelerators. This study used a Xilinx VC709 processor for comparison and employed Virtex™ 7VC707 (28nm) technology.
\subsection{ODE-based acceleration}

In 2024, a hybrid approach was proposed for the acceleration of the transformer networks by Okubo et al\cite{hw_fpga_ode}. The proposed scheme uses ResNet as a backbone architecture and replaces a part of its convolution layers with an MHSA (Multi-Head Self-Attention) mechanism. Using this approach they manage to significantly reduce the parameter size of such models by using Neural
ODE (Ordinary Differential Equation) as a backbone architecture instead of ResNet. The proposed hybrid model reduces the parameter size by 94.6\% compared to the CNN-based ones without degrading the accuracy.

The performance evaluation on a Xilinx Zynq UltraScale+ MPSoC platform shows that the proposed FPGA implementation achieves 12.8× speedup and 9.2× energy efficiency compared to an ARM Cortex-A53 CPU implementation.

\subsection{Beta}
In 2024, Yuhao Ji presented a Binary Transformer Accelerator (BETA) that achieves high performance and flexibility \cite{2024beta}. This is accomplished through a computational flow subtraction method aimed at optimizing QMMs. The QMM is a programmable machine that can support a wide range of precision while providing high parallelism, speed, and energy efficiency. Various experiments compared BETA with previous FPGA accelerators, concluding that energy efficiency continuously improves. While performance speed compared to other CPUs and GPUs is not mentioned, the energy efficiency is reported to be 22x better. The study used the RTX3090 and ZCU102 (16nm) technology, with BETA achieving an acceleration rate and energy efficiency rate of 1436 GOPs and 174 GOPs/W, respectively.

\subsection{Me-Vit}
In 2024, Kyle Marino, Pengmiao Zhang, and Viktor K. Prasanna introduced Me-ViT \cite{marino2024mevit}, a memory-efficient Vision Transformer design that outperforms traditional ViT accelerators on FPGA in speed and energy efficiency. Me-ViT combines Self-Attention and Multi-Layer Perceptron blocks, reducing data transfers and intermediate writes by loading weights only once. Its Memory-Efficient Processing Element (ME-PE) minimizes data movement and computation interruptions. Using systolic arrays for matrix multiplication, Me-ViT optimizes memory access, providing scalable, high-performance solutions for vision tasks on FPGA. Compared to CPUs and GPUs, Me-ViT is 5.1x faster and 4x more energy-efficient, achieving an acceleration rate of 2,682 GOPs. The study uses Nvidia TITAN RTX GPU and Alveo U200 (16nm) technology for comparison

\subsection{TransAxx}

In 2024, Dimitrios Danopoulos, Georgios Zervakis, and Dimitrios Soudris introduced TransAxx \cite{danopoulos2024transaxx}, a framework aimed at enhancing the efficiency of Vision Transformer (ViT) models through approximation computing. It includes a PyTorch-based system that supports continuous approximation computing and assesses its effects on ViT models. The technique involves studying the sensitivity of transformers to approximate multipliers, fine-tuning for accuracy, and using the Monte Carlo Tree Search (MCTS) algorithm to create approximate accelerators. Key techniques for accuracy improvement include quantization, pre-calibration training, and adaptive retraining. The framework reduces computational complexity and memory demands while balancing speed and energy efficiency. TransAxx provides a comprehensive approach for optimizing ViT models, enabling professionals to improve performance with limited resources through methods like quantization, calibration, and retraining.

\subsection{Ikumo Okubo}
In 2024, Ikumi Okubo introduced a cost-effective FPGA implementation of the Tiny Transformer model utilizing a Neural Ordinary Differential Equation (Neural ODE) technique \cite{okubo2024costefficient}. This method uses fewer parameters and less memory compared to ResNet-based deep models, making it suitable for resource-constrained devices. The model features ODEBlocks that reuse parameters, a learned relative positional encoding, and quantization to n-bit integers using LLTs. It also incorporates Depth-wise Separable Convolution (DSC) and Multi-Head Self-Attention (MHSA), forming a hybrid architecture. This approach is highly memory-efficient and significantly improves speed and energy efficiency, being 12.8x faster and 9.2x more energy-efficient than other models, and is compared to the ARM Cortex-A53 CPU using ZCU102 (16nm) technology.

\subsection{SSR}
In 2024, Jinming Zhuang presented SSR\cite{Zhuang_2024} as a unique architecture emphasizing the balance between latency and performance in accelerating transformers. It employs various elements such as FPGA and examines the trade-off between latency and performance for different models, achieving performance and energy efficiency increases. The method used is matrix multiplication, which controls the data communication between accelerators and seeks ways to improve performance. SSR provides open-source tools for reproducing results and can optimize communication between accelerators, reducing data transmission costs. Compared to other CPUs and GPUs, SSR is approximately 36x faster and 21x more energy-efficient than previous accelerators. This study utilizes the Nvidia A10G GPU and VCK190 (7nm) technology.

\section{CPU and GPU-based Accelerators}

\subsection{TurboTransformer}

In 2021, Jiarui Fang and Yang Yu introduced the TurboTransformers accelerator \cite{TurboTransformer}, a technique for efficiently serving Transformer models on GPUs for variable-length inputs. They addressed the challenges of padding smaller sequences to match the length of the longest sequence in a batch. By using dynamic programming to solve the optimization issue, they increased the response rate by 35 \% compared to not using batching.

To reduce memory size, TurboTransformers introduces a variable-length allocator that employs a segment-based memory management technique and a space reuse mechanism in the computation graph, reducing memory usage by 50 per cent compared to a reference allocator. Testing the system with various Transformer models, including BERT and Albert, the authors found that TurboTransformers outperformed PyTorch and ONNXRuntime in latency and performance for variable-length inputs, being 2.8x faster

\subsection{Jaewan Choi}

In 2022, researcher Jaewan Choi presented the study titled "Accelerating Transformer Networks through Rewiring of Softmax Layers"\cite{jaewan_choi}, which provides a method to accelerate the Softmax layer in transformer networks. The research introduces a rewiring technique to speed up the Softmax layer in transformer networks, which has become increasingly important as transformer models process longer sequences to improve accuracy rates. The proposed technique divides the Softmax layer into several sub-layers, changes the data access pattern, and then merges the disassembled Softmax sub-layers with the subsequent and preceding processes. This method accelerates the inference of BERT, GPT-Neo, BigBird, and Longformer on a current GPU by up to 1.25x, 1.12x, 1.57x, and 1.65x respectively, significantly reducing off-chip memory traffic.

\subsection{SoftMax}

In 2022, Choi et al. presented a novel framework for acceleration of transformer networks through Recomposing Softmax Layers\cite{2022_softmax}. The softmax layer normalizes the elements of the attention matrix to values between 0 and 1. This operation is conducted along the row vector of the attention matrix. Based on the profiling, the softmax layer in the scaled dot-product attention (SDA) block uses 36\%, 18\%, 40\%, and 42\% of the total execution time of BERT, GPT-Neo, BigBird, and Longformer, respectively.

Softmax recomposition achieves up to 1.25×, 1.12×, 1.57×, and 1.65× speedups in inferring BERT, GPT-Neo, BigBird, and Longformer on a A100 GPU by significantly reducing the off-chip memory traffic.

\subsection{LightSeq2}

In 2022, Wang et al. proposed a series of GPU optimizations to accelerate the training for a general family of Transformer models on GPUs called LightSeq2 \cite{2022_lightseq2}. 

LightSeq2 proposes 3 techniques for the acceleration of the training of transformer networks. 
Firstly, to all types of transformers, LightSeq2 uses fused kernel operators for both encoder and decoder
layers. Adjacent fine-grained element-wise kernels are fused into one coarse-grained kernel,
resulting in fewer kernel launches and intermediate results. For example, the last kernel of the self-attention layer implements bias adding, dropout, and residual kernels with only one kernel launch.

The performance evaluation shows that LightSeq2 is consistently faster (1.4-3.5×) than previous systems on different GPUs and it can achieve up to 3x speedup on large public datasets.

\subsection{Simplified Transformer Networks}

In 2023, He and Hofmann \cite{he2023simplifying} have also proposed a novel framework to accelerate transformer networks in GPUs by simplified transformers without compromising convergence properties and downstream task performance.

Based on the performance evaluation both on autoregressive decoder-only and BERT encoder-only models, the simplified transformers emulate the per-update training speed and performance of standard transformers, while enjoying 15\% faster training throughput in GPUs, and using 15\% fewer parameters. 

\subsection{LLMA}
In 2023, Nan Yang introduced LLMA\cite{yang2023inference}, an accelerator for large language models (LLMs) that enhances inference speed through interaction with reference data. This method uses a reference-based decoding mechanism to select and process tokens efficiently, enabling parallel execution on GPUs without needing new models. LLMA is easy to implement and deploy, providing over twice the speed for various model sizes using the Nvidia 32G V100 GPU.

\subsection{FlexGen}

In 2023, researchers introduced FlexGen, a high-throughput system for generating large language models (LLMs) designed for latency processing in resource-limited environments. FlexGen generates 32 tokens per prompt and evaluates throughput by the number of tokens generated divided by adaptation and decoding time. Compared to DeepSpeed ZeRO-Inference and Hugging Face Accelerate, FlexGen provides 40x more throughput with the same latency using an Nvidia T4 (16GB) GPU. Built on PyTorch, FlexGen utilizes multiple CUDA streams and CPU threads for I/O combination, significantly increasing performance through CPU computation and result overlapping.

\subsection{vLLMs}

In 2023, researchers introduced the vLLMs model to address efficient memory management for large language models (LLMs), which have high memory requirements \cite{vLLMs}. They proposed a strategy called PagedAttention, which divides key-value attention into fixed-size blocks and uses paging to maintain them. This approach enhances memory efficiency and reduces the memory footprint of LLMs. The vLLM architecture leverages PagedAttention to manage memory effectively, particularly in beam search scenarios with a fixed number of candidates. The model supports mixed decoding approaches with various sharing and memory access patterns, using a mapping layer to convert logical blocks to physical blocks, further optimizing memory usage and reducing the overall memory footprint of LLMs.

\subsection{Alisa}

In 2024, researchers introduced the ALISA model \cite{zhao2024alisa}, aimed at accelerating large language models (LLMs) through sparse window attention (SWA) and dynamic scheduling. This approach addresses the limitations of existing optimizations in maintaining competitive accuracy. SWA creates sparse patterns that are both locally static and globally dynamic, preserving the sequential semantics of language while capturing its dynamic evolution. Dynamic scheduling further enhances performance by balancing memory access and token processing. By integrating SWA, dynamic scheduling, and KV compression, ALISA significantly reduces the memory footprint of KV stores. The study demonstrates that ALISA outperforms previous methods in accuracy and performance, with comparisons across three families of open-source LLM models.

\section{ASIC Accelerators}

\subsection{A3}

One of the early research on the acceleration of transformer networks was proposed in 2020 by Hma et al. called A3 \cite{ham2020a3}. The paper proposes a hardware accelerator for attention mechanisms in NNs, that not only focused
on the efficient implementation of the attention mechanism in hardware but also on reducing the amount of computation in attention mechanism through algorithmic optimization and approximation. It presents an approximate candidate selection mechanism to reduce the number of search targets, and thus the amount of computation.

The proposed scheme has not been implemented on FPGA but is has been implemented on a cycle-accurate Verilog design  targeting a TSMC 40nm ASIC clocked at 1GHz. Based on the performance evaluation, the proposed scheme can achieve up to 7x speedup compared to a Intel Gold 6128 CPU implementation and up to 11x better energy efficiency against versus a CPU implementation.

\subsection{ELSA}

In 2021, Ham et al. presented a hardware-software Co-design approach for the acceleration of transformer networks called Elsa\cite{2021_elsa}. 

Based on the fact that irrelevant relations can be effectively filtered out by computing approximate similarity,
ELSA substantially reduces computational waste in a selfattention operation. Unlike conventional hardware such as CPU or GPUs, that cannot benefit from approximation, ELSA propose a specialized hardware that directly translates this reduction to further improve performance and energy efficiency. 

They evaluate several representative self-attention-oriented NN models to demonstrate the effectiveness of the ELSA. For performance evaluation, they implemented a custom simulator for ELSA targeting a 40nm ASIC clocked at 1GHz. ELSA-moderate achieves up to 157x speedup compared to GPUs and two orders of magnitude improvements in energy efficiency over the GPU for the self-attention computation. 

\subsection{SpAtten}

In 2021, Want et al. presented a framework for the acceleration for large language models called Spatten. \cite{2021_spatten}

SpAtten propose a novel scheme for the acceleration of NLP using three algorithmic optimizations: cascade token pruning, cascade head pruning and progressive quantization to reduce computation and memory access.

The proposed scheme has been implemented on a cycle-accurate design using SpinalHDL and mapped to ASIC using a 40nm TSMC library. SpAtten  and achieves 162x, and 347x speedup over a GPU (TITAN Xp), and a Xeon CPU, respectively. In terms of energy efficiency SpAtten achieves 1193x and 4059x energy savings compared to GPU and CPU.

\subsection{Sanger}

Lu at el. presented in 2021 another novel approach for the acceleration of transformer networks called Sanger\cite{2021_sanger}. 

Sanger accelerates the sparse attention models by combining dynamic sparsity patterns and reconfigurable architecture.
The software part provides sparsity patterns, which can achieve high performance and a balanced workload. The architecture is designed with reconfigurability to support the dynamic characteristics of sparsity, which helps to improve the compression ratio.

To allow more flexibility in the sparsity patterns, Sanger proposed a reconfigurable systolic array based on this dataflow. Sanger was implemented in Chisel hardware that was translated to Verilog RTL. The design was targeting an ASIC using the UMC 55nm technology clocked at 500MHz.

\subsection{SALO}
In 2022, Guan Shen presents a spatial accelerator to improve transformer performance for long sequences by using hybrid sparse attention patterns \cite{salo}. This model addresses computational and memory challenges, achieving 25.5x faster performance and 336.1x better energy efficiency compared to a Nvidia GTX 1080Ti with FreePDK (45nm) technology. It demonstrates enhanced attention processing through efficient hardware and data strategies

\subsection{AccelTran}
In 2020, Shikhar Tuli introduced AccelTran \cite{AccelTran}, an accelerator architecture to improve transformer model efficiency in natural language processing. AccelTran uses matrix compression and data flow strategies to enhance energy efficiency. It has two versions: AccelTran-Edge for low-power portable devices and AccelTran-Server for high-throughput cloud applications. AccelTran-Edge outperforms Raspberry Pi in throughput and power consumption, while AccelTran-Server offers 5.7x higher throughput and 3.7x lower energy use compared to the Energon model. The acceleration rates are 372,000 GOPs for AccelTran-Server and 7,520 GOPs for AccelTran-Edge

\subsection{DTQAtten}
In 2022, Tao Yang introduced DTQAtten \cite{DTQAtten1}, a technique to enhance NLP model efficiency by combining dynamic quantization and specialized hardware. DTQAtten reduces memory usage, inference latency, and energy consumption in models like BERT and GPT-2. It is 16.4x faster and 3.8x more energy-efficient than the SpAtten model on Nvidia Titan Xp using TSMC (40nm) technology, achieving 952 GOPs throughput and 1298.4 GOPs/W energy efficiency. This makes DTQAtten significantly superior to advanced accelerators such as A3 and SpAtten.

\subsection{Energon}

In 2023, Zhou et al. presented a algorithm-architecture co-design approach that accelerates various transformers using dynamic sparse attention, called Energon \cite{2023_energon}. Energon proposes a mix-precision multi-round filtering (MP-MRF) algorithm to dynamically identify query-key pairs at runtime. 

Energon adopts low bit-width in each filtering round and only the finally selected pairs are used for high-precision tensors in the attention stage to reduce overall complexity. By this means, they manage to reduce by 4× to 8× the computation cost with negligible accuracy loss. 

Energon is implemented as a co-processor targeting a 45nm ASIC library. 
Based on the performance evaluation it is shown Energon achieves 168× and 8.7× speedup and up to 10000× and 1000× energy reduction over Intel Xeon 5220 CPU and NVIDIA V100 GPU, respectively. 

\subsection{H3D Transformer}

In 2024, a new 3D heterogeneous accelerator design was proposed for transformer models \cite{h3d}. It combines compute-in-memory (CIM) and digital tensor processing units (TPUs) to address chip area and energy consumption issues. The design uses 22nm FeFET digital CIM chips for high-density on-chip memory and processes MatMul tasks efficiently. It achieves 10 TOPS/W (10,000 GOPs/W) for BERT and GPT-2 models, which is 2.6x to 3.1x better than 7nm TPU and FeFET memory baselines.

\subsection{SALO2}
In 2024, Jieru Zhao introduced SALO2\cite{salo2}, an enhanced framework for efficient attention computation in both static and dynamic sparsity scenarios. Combining software optimizations with hardware accelerators and pattern matching units, SALO2 improves performance and flexibility for various applications. It uses sparse attention algorithms and data reorganization techniques. SALO2 is about 25x faster and 70x more energy-efficient than previous models, using Nvidia RTX4090 with FreePDK (45nm) technology as a benchmark.

\section{In-Memory Hardware Accelerators}

\subsection{ATT}

In 2020, Guo et al. presented another approach for the acceleration of attention-based accelerators called ATT \cite{2020_att} based on resistive RAM. ATT is based on crossbar-based resistive RAM that can eliminates weight movement between memory and processing units with a dedicated pipeline design for Attention-based Neural Networks. The proposed scheme consists of several modules.

The proposed scheme has been simulated using CACTI 7.0 at 32 nm to model the power and area of the SRAM buffer and the Mask Cache. Based on the performance evaluation, ATT can achieve 202x speedup compared to NVIDIA GTX 1080 Ti GPU.

\subsection{ReTransformer}

In 2020, Yang et al. proposed an in-memory framework for the acceleration of transformers called ReTransformer \cite{hw_retransformer}. ReTransformer is a ReRAM-based In-Memory architecture for Transformer acceleration that is not only accelerate the scaled dot-product attention of Transformer using ReRAM-based In-Memory architecture but also eliminate some data dependency by avoiding writing the intermediate results using the proposed matrix decomposition technique. Furthermore, ReTransformer proposes a new sub-matrix pipeline design for multi-head self-attention.

The performance evaluation shows that compared to GPU, ReTransformer can achieve up to 23.21× speedup while the corresponding overall power is reduced by 1086×.

\subsection{iMCAT}

In 2021, Laguna et al. presented a novel in-memory architecture for the acceleration of transformer networks for long sentences called iMCAT \cite{hw_inmemory}. The proposed framework uses a combination of XBars and CAMs to accelerate transformer networks. The acceleration of transformer networks is achieved by combining several techniques such as computing in-memory, thus minimizing the memory transfer overhead, caching reusable parameters to reduce the number of operations, exploiting the available parallelism in the attention mechanism, and finally using locality sensitive hashing to filter the number of sequence elements by their importance. 

The performance evaluation shows that this approach achieves a 200x speedup and 41x energy improvement for a sequence length of 4098.

\subsection{TransPIM}

In 2023, researchers introduced TransPIM\cite{transpim} to enhance Transformer operations through combined software and hardware design. By adding auxiliary computation units (ACUs) to memory, it improves vector reduction and Softmax algorithms. TransPIM increases speed and energy efficiency by optimizing parallelism, reducing latency, and energy use up to 10.8 and 5.7x, respectively. It also uses token-based data division to boost memory-level parallelism and reduce data transfer costs. This architecture achieves an acceleration rate of 734 GOPs, significantly enhancing Transformer model performance

\subsection{iMTransformer}
In 2022, Ann Franchesca Laguna introduced iMTransformer, an architecture for hardware-software co-design in memory systems. It leverages Multi-Head Attention (MHA) in transformer networks for parallel processing and details encoder and decoder layers' roles. The study covers masked, bidirectional, and encoder-decoder MHA types and explores sparse attention techniques to improve efficiency. iMTransformer enhances parallelism and reduces energy use, achieving 11x faster speed and 12.6x better energy efficiency compared to Nvidia Titan RTX GPU using CAMS (12nm) technology.

\subsection{X-Former}

In 2023, Sridharan et al. presented a novel in-memory hardware acceleration to speedup transformer networks called X-Former\cite{hw_xformer}. X-Former is a hybrid spatial in-memory hardware accelerator that consists of both NVM and CMOS processing elements to execute transformer workloads efficiently. 

X-Former is composed primarily of a Projection Engine with NVM processing tiles for executing MV MStatic operations and an Attention Engine with CMOS processing tiles for executing MV MDynamic operations. The main difference compared to other in-memory architectures is that the weights of all the layers are stored in the Projection Engine to prevent reprogramming the NVM tiles, while the Attention Engine is optimized to only process the largest self-attention layer due to area constraints. 

Based on the performance evaluation it is shown that X-Former achieves up to 85x and 7.5x improvements in latency and energy over a NVIDIA GeForce GTX 1060 GPU and upto 10.7x and 4.6x improvements in latency and energy over a state-of-the-art in-memory NVM accelerator.

\subsection{TranCIM}
In 2023, researchers introduced TranCIM\cite{trancim}, an innovative design to improve transformer model efficiency in NLP, computer vision, and bioinformatics. TranCIM addresses data transfer and processing challenges with a fully digital CIM accelerator that enhances memory access and computations for attention and fully connected layers. It uses a pipeline function with bitline transfer architecture for efficient matrix computation. Compared to other models and Nvidia Jetson Nano, TranCIM is 16.9x faster and 1.6x more energy-efficient using CMOS (28nm) technology.

\subsection{H3DATTEN}
n 2023, researchers introduced H3DAtten\cite{h3datten}, an architecture enhancing vision transformer efficiency. Combining analog (ACIM) and digital (DCIM) in-memory computing, H3DAtten uses the Swin transformer for accurate image recognition and object detection. It processes input features in variable-sized windows for multi-scale feature extraction. ACIM converts analog data to digital bits, while DCIM uses SRAM for MatMul operations, reducing energy and latency. H3DAtten achieves 1600 GOPs acceleration and 7100 GOPs/W energy efficiency, outperforming existing hardware accelerators.

\subsection{PRIMATE}
In 2024, the Primate\cite{primate} framework was introduced to accelerate transformer models using dynamic token pruning and PIM (Processing-In-Memory) technology. Primate offers higher capacity, greater bandwidth, lower latency, and stable computations. It uses a pipeline strategy to maximize parallelism and efficiency. Primate improves performance by 30.6x, space efficiency by 29.5x, and energy efficiency by 4.3x compared to the TransPIM model based on Nvidia RTX 2080Ti.

\subsection{HARSEA}
In 2024, researchers introduced HARDSEA\cite{hardsea}, an accelerator architecture for transformer models focusing on the self-attention mechanism. HARDSEA uses a hybrid analog-digital approach with sparse self-attention and digital SRAM-CIM for in-memory computation, reducing costs by leveraging sparsity. It shows excellent performance with transformer models, achieving 28.5x better acceleration and 1,894.3x  better energy efficiency compared to Nvidia RTX 3090. HARDSEA has an acceleration rate of 921.6 GOPs and an energy efficiency rate of 943.7 GOPs/W.

\section{Quantitative comparison}

Table \ref{tab:list-all} shows all of the hardware-based accelerators that have been proposed and the main features for each accelerator. Each row presents the name of the accelerator, the type of the accelerator (FPGA/ASIC/In-memory), the performance and the energy-efficiency. In some cases, the papers were also mentioning the speedup when the proposed architecture is compared against a CPU and a GPU. However, since the baseline comparison for each architecture was different we present only the absolute performance and energy efficiency and not the speedup for each architecture.

\begin{table*}
 \centering
  \caption{LLM-Transformer Accelerators}
  \label{tab:list-all}
  \begin{tabular}{ccccc}
\hline
\makecell {Year\\} & \makecell{Framework\\}          & \makecell{Technology\\}    & \makecell{Performance \\(GOPs)}  & \makecell{Energy efficiency \\  (GOPs/W)} \\  
\hline

2020 & FTRANS\cite{2020_ftrans}      & FPGA VCU118   & 170   &  6.8   \\ 
2022 & Via\cite{2021_via}              & FPGA Alveo U50  & 309.6 & 7.9 \\ 
2022 & STA-4\cite{sta}                & FPGA Arria 10SX660  & 392.9 & -- \\ 
2002 & STA-8\cite{sta}               & FPGA Arria 10SX660  & 523.8 & 41.2 \\
2023 & Zhongyo Zhao \cite{}      & FPGA Virtex 7VC707  & 728.3 & 58.3 \\ 
2023 & Swin-T \cite{swin}            & FPGA XCZU19EG  & 431.2 & -- \\ 
2023 & Swin-B\cite{swin}             & FPGA XCZU19EG  & 403.5 & -- \\ 
2023 & Swin-S\cite{swin}             & FPGA XCZU19EG  & 436.4 & -- \\ 

2024 & BETA\cite{2024beta}      & FPGA  ZCU102 &1436 & 174 \\ 
2024 & Me-ViT\cite{marino2024mevit} & FPGA Alveo U200 & 2682 &-- \\ 

2020 & A3\cite{ham2020a3}            & ASIC 40nm     & 221       & 269 \\
2021 & SpAtten\cite{2021_spatten}    & ASIC 40nm     & 360 & 382 \\
2021 & Sanger\cite{2021_sanger}      & ASIC 55nm     & 529 & ---    \\
2022 & AccelTran (edge)\cite{AccelTran}    & ASIC 14nm     & 7520 & -- \\
2022 & AccelTran (server)\cite{AccelTran}    & ASIC 14nm     & 372000 & -- \\
2024 & H3D Transformer\cite{h3d}    & ASIC 22nm     & 1600 &  -- \\

2020 & ReTransformer\cite{hw_retransformer} & In-memory & 81.9    & 467.7 \\ 
2022 & TransPiM\cite{transpim}       & In-memory     & 734     &  --\\
2023 & X-Former\cite{hw_xformer}     & In-memory     & --       & 13440 \\ 
2023 & H3DAtten\cite{h3datten}     & In-memory     & 1600       & 7100   \\ 
2023 & TranCIM\cite{trancim}     & In-memory     & --       & 20500    \\ 
2024 & Hardsea\cite{hardsea}     & In-memory     & 921.6       & 943.7   \\ 
\hline
  \end{tabular}
\end{table*}

\subsection{Quantitative Comparison on performance}

\begin{figure*}
    \centering
    \includegraphics[width=0.8\linewidth]{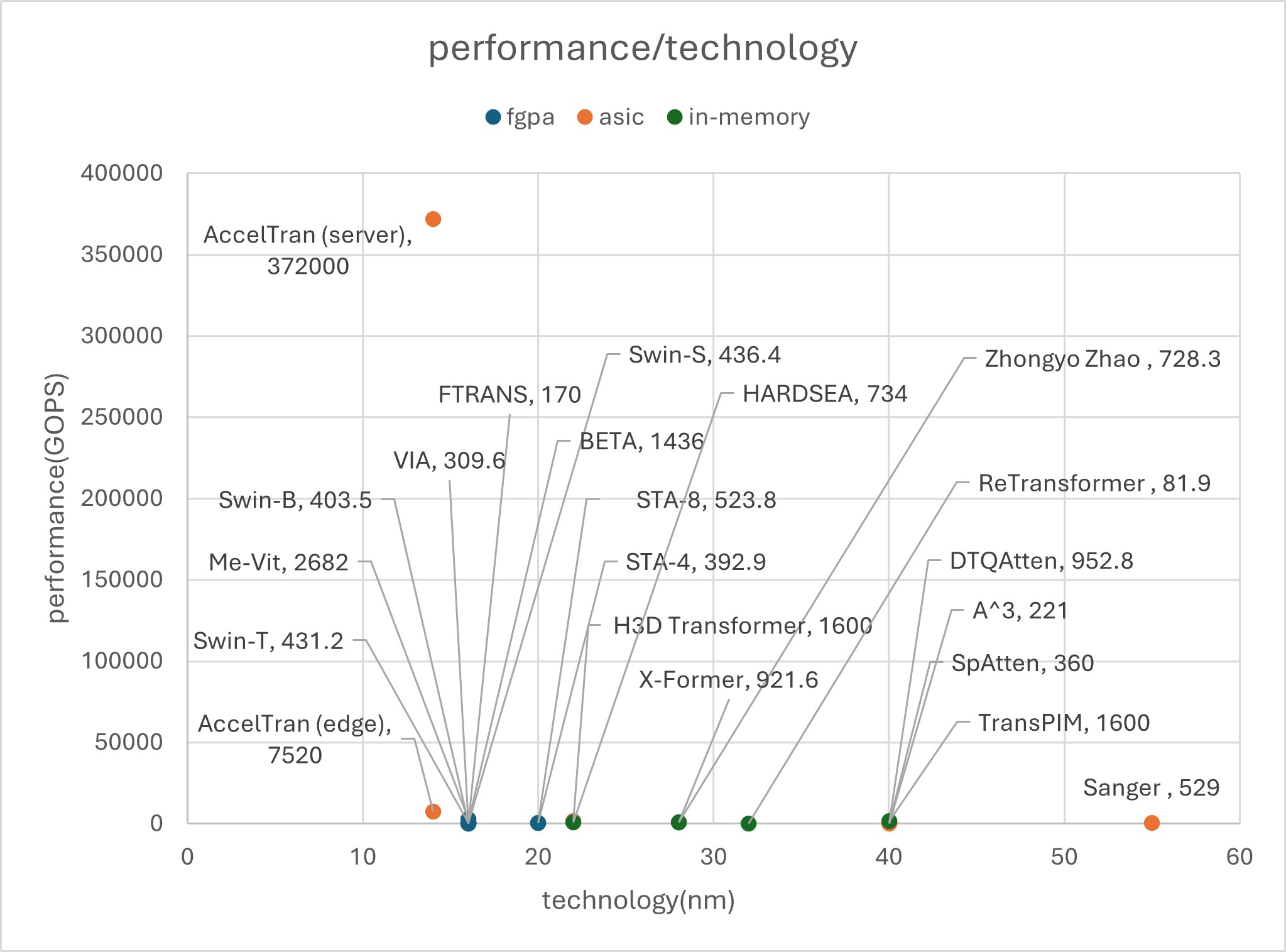}
    \caption{Performance per process technology}
    \label{fig:performence-nm}
\end{figure*}

\begin{figure*}
    \centering
    \includegraphics[width=0.8\linewidth]{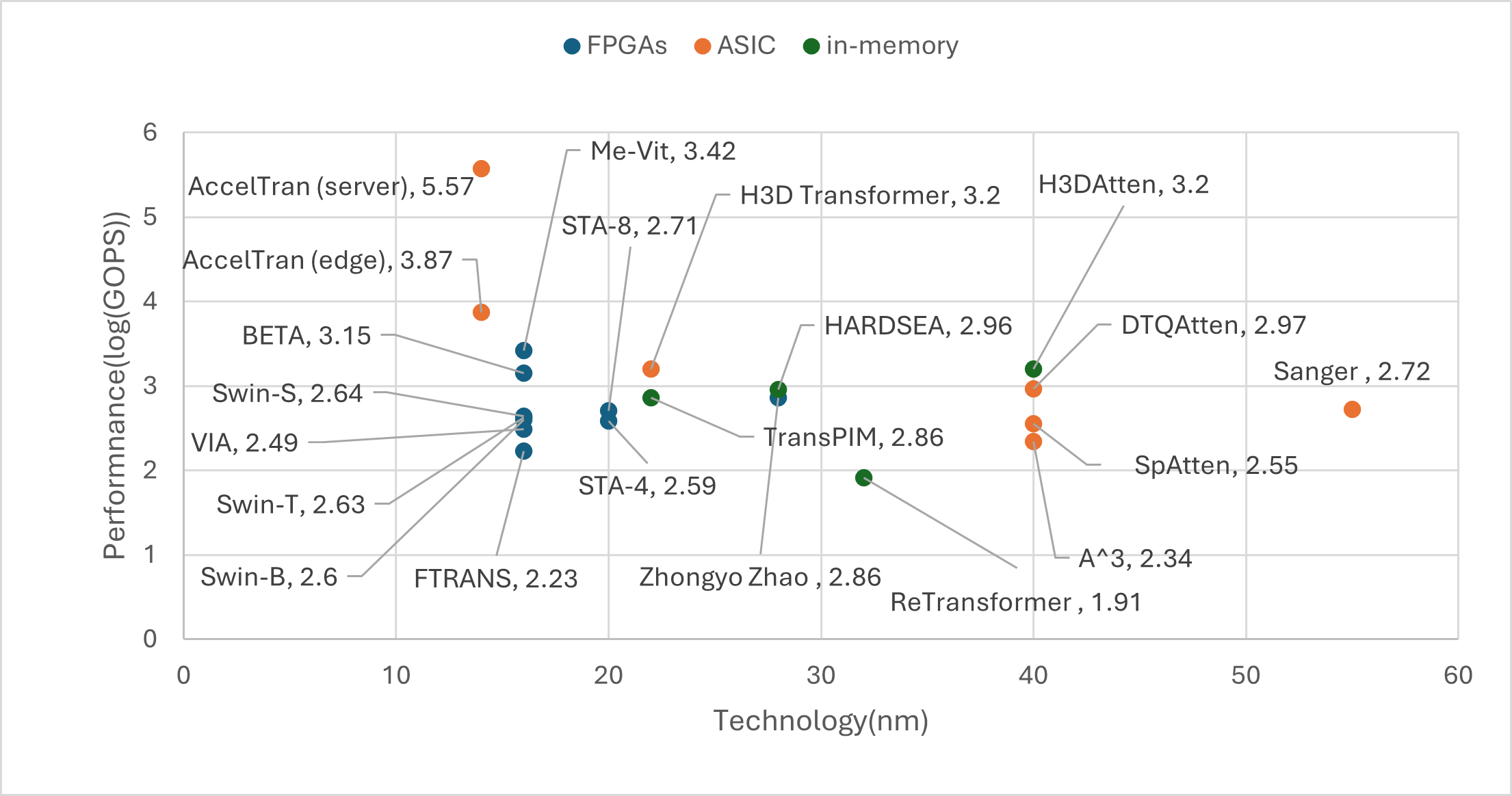}
    \caption{Performance (Log) per process technology}
    \label{fig:performence-nm-log}
\end{figure*}

Figure \ref{fig:performence-nm} shows the performance of each accelerator based on the process technology while Figure \ref{fig:performence-nm-log} shows the performance in logarithmic scale for better visibility.

As it is shown in the table and on the figures, the highest performance is achieved by the AccelTran (server) architecture using 14nm process technology with 372,000 GOPs, while the lowest is by the ReTransformer model. Additionally, it is observed that models within the same technology, such as ViA, Me-ViT, Ftrans, and others, do not have similar performance. However, for the accelerators that do not use the same process technology it is hard to make a fair comparison as the process technology is affects significantly the performance of a hardware accelerator. 
 
\subsection{Comparison of Energy efficiency vs process technology}

\begin{figure*}
    \centering
    \includegraphics[width=0.8\linewidth]{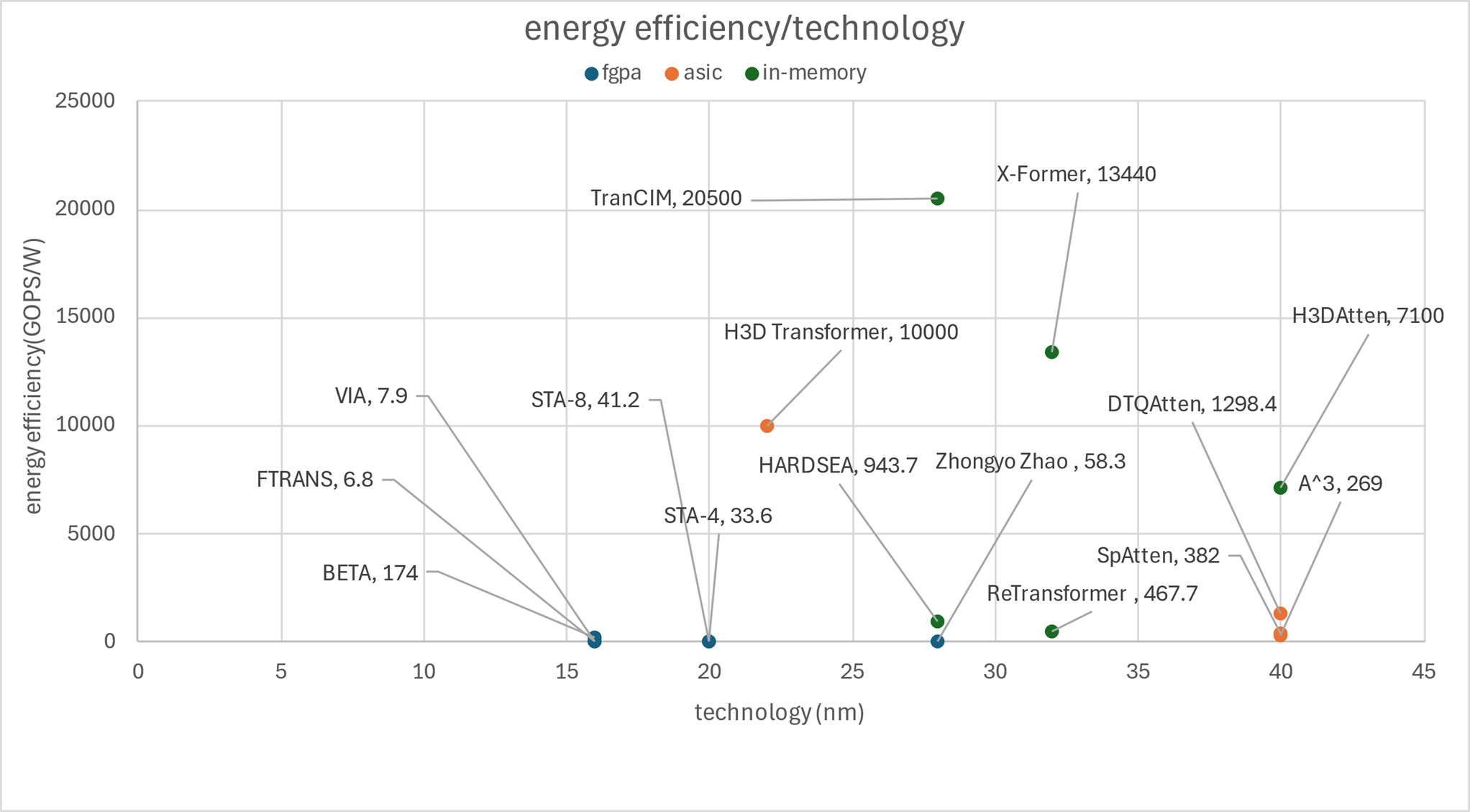}
    \caption{Performance per process technology}
    \label{fig:energy-nm}
\end{figure*}

\begin{figure*}
    \centering
    \includegraphics[width=0.8\linewidth]{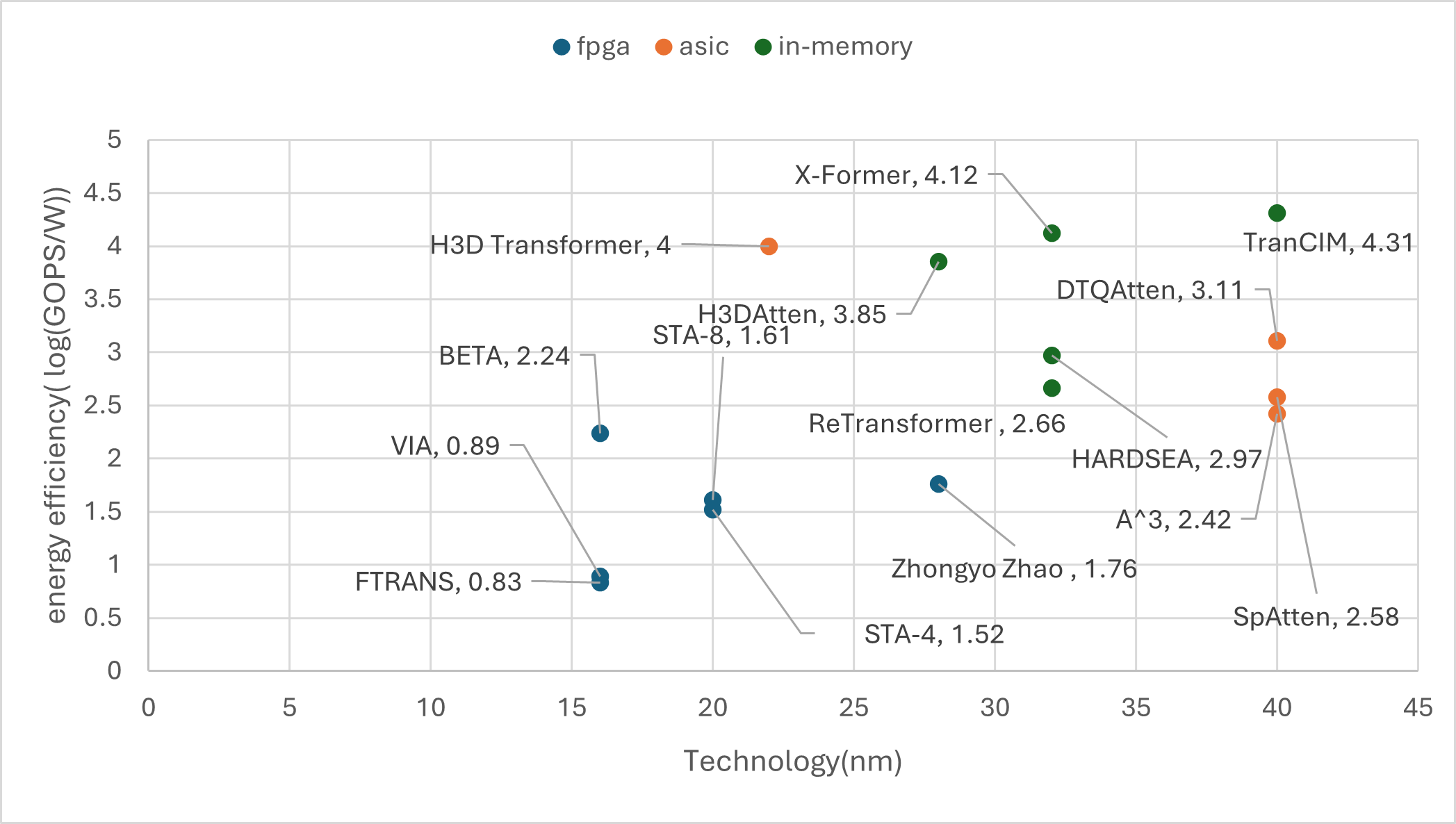}
    \caption{Energy efficiency (log) per process technology}
    \label{fig:energy-nm-log}
\end{figure*}

Figure \ref{fig:energy-nm} shows the energy efficiency in terms of GOPs/W for most of the hardware accelerators while Figure \ref{fig:energy-nm-log} shows the energy efficiancy in logarithmic scale for better visibility. Since many proposed architectures do not measure the energy efficiancy, we list only the accelerators that present the energy efficiency. Again, many accelerators are based on different process technology and therefore it is hard to make a fair comparison. 

The models using primarily memory (In-Memory Accelerators) have better energy efficiency. This happens because they have reduced data movement, as this specific architecture allows data to be processed directly in memory, without being transferred from memory to the central processing unit (CPU).

\subsection{Comparison of acceleration in 16nm technology}
It is important to note that there is no easy way to compare acceleration and energy efficiency when the existing models have different characteristics and especially when they use different process technology that affects significantly the performance of the hardware accelerators. To make a fair comparison of the hardware accelerators we extrapolate the performance and the energy efficiency on the same process technology. As the in-memory accelerators the performance is not based only on the process technology, the extrapolation is performed only on the FPGA and ASIC accelerators where the process technology affects significantly the performance of the systems. 

Based on the article by Aaron Stillmaker and B.Baas titled "Scaling equations for the accurate prediction of CMOS device performance from 180 nm to 7nm,"\cite{key} we extrapolated the performance and the energy efficiency on a 16nm technology to make a fair comparison of acceleration and energy efficiency. Based on the tables and equations provided by the paper, we made the extrapolation of the features on the same technology. 

Tables \ref{tab:extrapolation} shows the extrapolated performance of the hardware accelerators for the 16nm process technology. 
\begin{table*}
 \centering
  \caption{FPGAs Accelerators- to 16nm process Technology}
  \label{tab:extrapolation}
  \begin{tabular}{cccc}

\hline
\makecell{Name \\ }         & \makecell{Technology \\ (nm)} & \makecell{New Technology \\nm} & \makecell{New Performance \\(GOPS)}  \\
  
\hline

FTRANS\cite{2020_ftrans}      & FPGA 16nm  & 16   & 170  \\
Via\cite{2021_via}            & FPGA 16nm  & 16   & 309.6 \\
STA-4\cite{sta}               & FPGA 20nm  & 16   & 586.68 \\
STA-8\cite{sta}               & FPGA 20nm  & 16   & 782.147 \\
Swin-T \cite{swin}            & FPGA 16nm  & 16   & 431.2 \\
Swin-B\cite{swin}             & FPGA 16nm  & 16   & 403.5 \\
Swin-S\cite{swin}             & FPGA 16nm  & 16   & 280.43 \\
BETA\cite{2024beta}           & FPGA 16nm  & 16   & 1436 \\
Me-ViT\cite{marino2024mevit}  & FPGA 16nm  & 16   & 2682 \\
AccelTran (edge)\cite{AccelTran}   & ASIC 14nm  & 16  &  7178 \\
AccelTran (server)\cite{AccelTran} & ASIC 14nm  & 16  &  363647\\

\hline

  \end{tabular}
\end{table*}

\begin{figure*}
    \centering
    \includegraphics[width=0.8\linewidth]{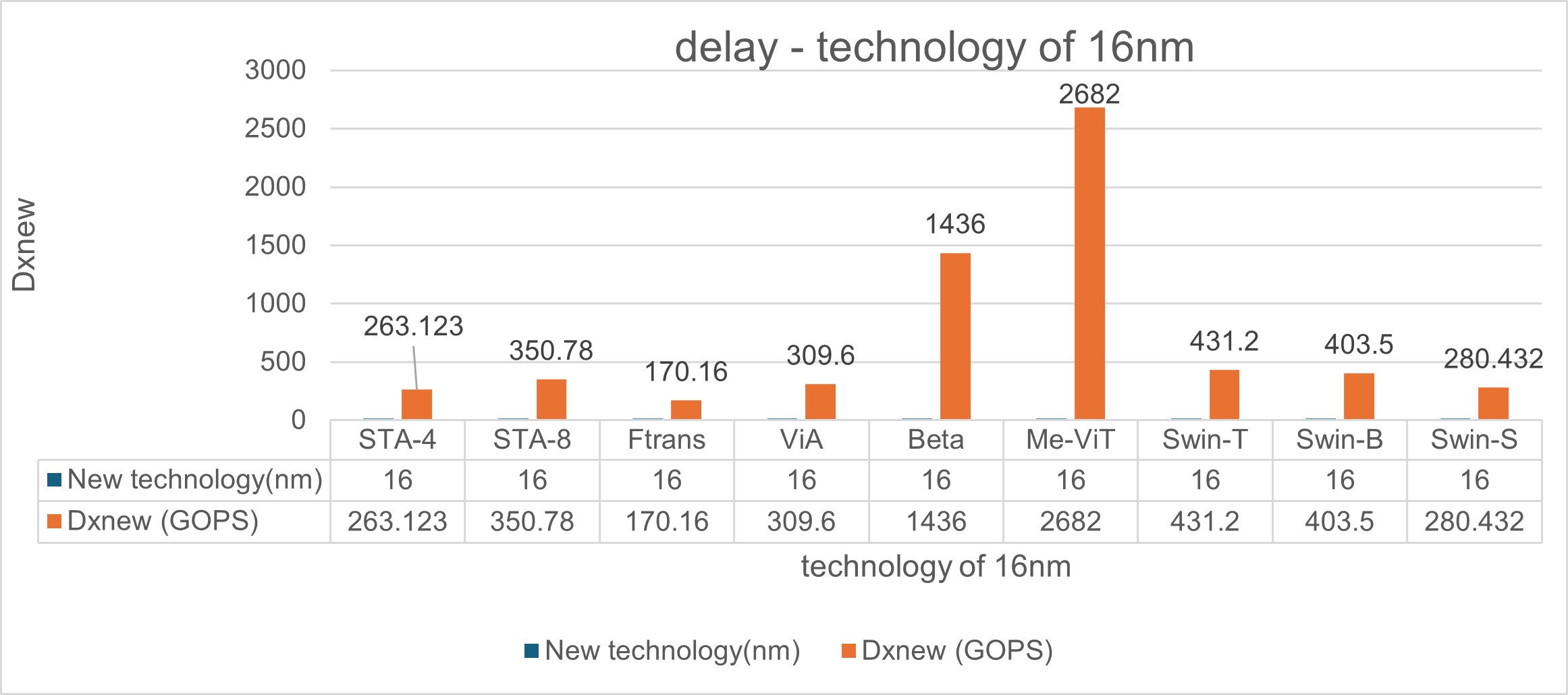}
    \caption{Extrapolated performance per process technology}
    \label{fig:performance-16nm}
\end{figure*}

Figure \ref{fig:performance-16nm} shows the absolute performance of the hardware accelerators when the performance is extrapolated on the same 16nm process technology. AccelTran achieves the highest performance (omitted in the diagram for better visibility of the rest of the results).

\subsection{Experimental extrapolation}

While the scaling equations provided by \cite{key} can help on the extrapolation of the performance on the same technology, we perfmed also an experimantal extrapolation for the architectures targetting FPGAs. 

The most computational intensive part of the LLMs is the matrix multiplication. Therefore we used a matrix multiplication IP core in VHDL\cite{vhdl} and we used Quartus EDA tool to evaluate the performance and accuracy of these accelerators. We tested the matrix multiplication code across various FPGA technologies, specifically 20nm, 28nm, 40nm, 55nm, 65nm, and 180nm, to verify the theoretical conversion to 16nm technology. The results of the matrix multiplication on several FPGA technologies can help extrapolate the results of the hardware accelerators on the same technology.

The FPGA device, the process technology and the results obtained from the matrix multiplications IP cores are listed in the Table \ref{tab:fpga-results}:

\begin{table*}
 \centering

  \caption{VHDL libraries}
  \label{tab:fpga-results}
  \begin{tabular}{ccc}
  \hline
Name         & Technology(nm)    & Fmax(MHz)  \\
  \hline
Ariia 10 & 20  & 182.02  \\
Starix V & 28  & 215.75\\
Starix IV  & 40  & 188.71\\
Arria II GX  & 40  & 133.87\\
Max 10  & 55  & 91.63\\
Cyclone IV & 65  & 84.04\\
Max V & 180  & 36.37\\
  \hline
  \end{tabular}
\end{table*}

\begin{figure*}
    \centering
    \includegraphics[width=0.8\linewidth]{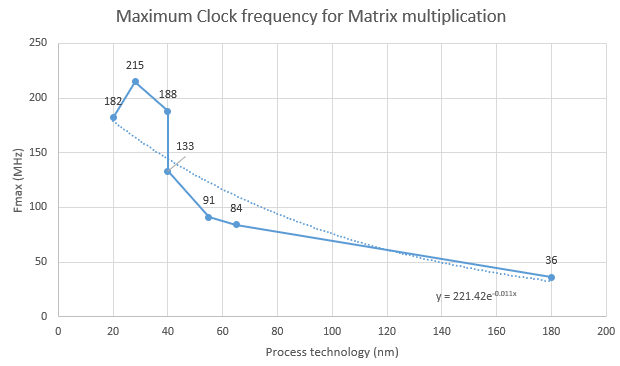}
    \caption{Maximum Clock frequency on each FPGA}
    \label{fig:fmax-nm}
\end{figure*}

Figure \ref{fig:fmax-nm} shows the maximum clock frequency achieved for each FPGA device and process technology for the matrix multiplications. The table below lists the FPGA accelerators, including the frequency at which they operated and the corresponding technology. The table shows the new frequency that is extrapolate on the 16nm process technology and the new performance achieved using the extrapolated performance. As the performance in FPGAs is depending on the maximum clock frequency, the extrapolated performance allows a fair comparison between architectures that were mapped on different process technologies. 

\begin{table*}
 \centering

  \caption{Extrapolated Frequency and performance on 16nm Process Technology}
  \label{tab:commands}
  \begin{tabular}{ccccc}
\hline
FPGAs & Frequency (MHz)  & Technology(nm) & New Frequency (MHz) & New Performance (GOPs)  \\
\hline

STA-8	        & 200	&   20	 & 202.8   & 3985\\
STA-4           & 200   &	20   & 202.8   & 531\\
ViA	            & 300   &   16   & 300	   & 309\\
Me-Vit          & 300	&   16	 & 300	   & 2682\\
Swin-T	        & 200	&   16	 & 200	   & 431\\
Swin-B	        & 200	&   16	 & 200	   & 403\\
Swin-S	        & 200	&   16	 & 200	   & 136\\
BETA	        & 190 	&   16   & 190	   & 1436\\
\hline
  \end{tabular}
\end{table*}

\section{Conclusions}

Large Language Models have been emerged as a promising and powerful technology for the science and the society in general. There are several research efforts on the acceleration of the LLMs. However, the use of different process technologies makes it hard to make a fair comparison. In this paper we extrapolated the performance and the energy-efficiency of the hardware accelerators to the same process technology using both a theoretical and an experimental methodology. Overall it seems that the in-memory architectures provide much better energy efficiency compared to other technologies. However, in terms of absolute performance (GOPs) it seems that accelerators targetting ASICs can provide much higher performance compared to other platforms.



\bibliographystyle{IEEEtran}

\bibliography{sample}

\begin{thebibliography}{10}
\providecommand{\url}[1]{#1}
\csname url@samestyle\endcsname
\providecommand{\newblock}{\relax}
\providecommand{\bibinfo}[2]{#2}
\providecommand{\BIBentrySTDinterwordspacing}{\spaceskip=0pt\relax}
\providecommand{\BIBentryALTinterwordstretchfactor}{4}
\providecommand{\BIBentryALTinterwordspacing}{\spaceskip=\fontdimen2\font plus
\BIBentryALTinterwordstretchfactor\fontdimen3\font minus \fontdimen4\font\relax}
\providecommand{\BIBforeignlanguage}[2]{{%
\expandafter\ifx\csname l@#1\endcsname\relax
\typeout{** WARNING: IEEEtran.bst: No hyphenation pattern has been}%
\typeout{** loaded for the language `#1'. Using the pattern for}%
\typeout{** the default language instead.}%
\else
\language=\csname l@#1\endcsname
\fi
#2}}
\providecommand{\BIBdecl}{\relax}
\BIBdecl

\bibitem{survey_transformer}
J.~Zhong, Z.~Liu, and X.~Chen, ``Transformer-based models and hardware acceleration analysis in autonomous driving: A survey,'' 2023.

\bibitem{hw_survey_huang}
\BIBentryALTinterwordspacing
S.~Huang, E.~Tang, S.~Li, X.~Ping, and R.~Chen, ``Hardware-friendly compression and hardware acceleration for transformer: A survey,'' \emph{Electronic Research Archive}, vol.~30, no.~10, pp. 3755--3785, 2022. [Online]. Available: \url{https://www.aimspress.com/article/doi/10.3934/era.2022192}
\BIBentrySTDinterwordspacing

\bibitem{2023comprehensive}
M.~Emani, S.~Foreman, V.~Sastry, Z.~Xie, S.~Raskar, W.~Arnold, R.~Thakur, V.~Vishwanath, and M.~E. Papka, ``A comprehensive performance study of large language models on novel ai accelerators,'' 2023.

\bibitem{2020_ftrans}
\BIBentryALTinterwordspacing
B.~Li, S.~Pandey, H.~Fang, Y.~Lyv, J.~Li, J.~Chen, M.~Xie, L.~Wan, H.~Liu, and C.~Ding, ``Ftrans: Energy-efficient acceleration of transformers using fpga,'' in \emph{Proceedings of the ACM/IEEE International Symposium on Low Power Electronics and Design}, ser. ISLPED '20.\hskip 1em plus 0.5em minus 0.4em\relax New York, NY, USA: Association for Computing Machinery, 2020, p. 175–180. [Online]. Available: \url{https://doi.org/10.1145/3370748.3406567}
\BIBentrySTDinterwordspacing

\bibitem{2020_multihead}
S.~Lu, M.~Wang, S.~Liang, J.~Lin, and Z.~Wang, ``Hardware accelerator for multi-head attention and position-wise feed-forward in the transformer,'' in \emph{2020 IEEE 33rd International System-on-Chip Conference (SOCC)}, 2020, pp. 84--89.

\bibitem{2021_FPGA_NPE}
\BIBentryALTinterwordspacing
H.~Khan, A.~Khan, Z.~Khan, L.~B. Huang, K.~Wang, and L.~He, ``Npe: An fpga-based overlay processor for natural language processing,'' ser. FPGA '21.\hskip 1em plus 0.5em minus 0.4em\relax New York, NY, USA: Association for Computing Machinery, 2021, p. 227. [Online]. Available: \url{https://doi.org/10.1145/3431920.3439477}
\BIBentrySTDinterwordspacing

\bibitem{2021_pruning}
H.~Peng, S.~Huang, T.~Geng, A.~Li, W.~Jiang, H.~Liu, S.~Wang, and C.~Ding, ``Accelerating transformer-based deep learning models on fpgas using column balanced block pruning,'' in \emph{2021 22nd International Symposium on Quality Electronic Design (ISQED)}, 2021, pp. 142--148.

\bibitem{column_balanced}
------, ``Accelerating transformer-based deep learning models on fpgas using column balanced block pruning,'' in \emph{2021 22nd International Symposium on Quality Electronic Design (ISQED)}, 2021, pp. 142--148.

\bibitem{2021_via}
T.~Wang, L.~Gong, C.~Wang, Y.~Yang, Y.~Gao, X.~Zhou, and H.~Chen, ``Via: A novel vision-transformer accelerator based on fpga,'' \emph{IEEE Transactions on Computer-Aided Design of Integrated Circuits and Systems}, vol.~41, no.~11, pp. 4088--4099, 2022.

\bibitem{2022_DFX}
S.~Hong, S.~Moon, J.~Kim, S.~Lee, M.~Kim, D.~Lee, and J.-Y. Kim, ``Dfx: A low-latency multi-fpga appliance for accelerating transformer-based text generation,'' 2022.

\bibitem{sta}
C.~Fang, S.~Guo, W.~Wu, J.~Lin, Z.~Wang, M.~K. Hsu, and L.~Liu, ``An efficient hardware accelerator for sparse transformer neural networks,'' in \emph{2022 IEEE International Symposium on Circuits and Systems (ISCAS)}, 2022, pp. 2670--2674.

\bibitem{2023_OPU}
Y.~Bai, H.~Zhou, K.~Zhao, J.~Chen, J.~Yu, and K.~Wang, ``Transformer-opu: An fpga-based overlay processor for transformer networks,'' in \emph{2023 IEEE 31st Annual International Symposium on Field-Programmable Custom Computing Machines (FCCM)}, 2023, pp. 221--221.

\bibitem{tzanos}
G.~Tzanos, C.~Kachris, and D.~Soudris, ``Hardware acceleration of transformer networks using fpgas,'' in \emph{2022 Panhellenic Conference on Electronics and Telecommunications (PACET)}, 2022, pp. 1--5.

\bibitem{hw_fpga_flexrun}
\BIBentryALTinterwordspacing
S.~Hur, S.~Na, D.~Kwon, J.~Kim, A.~Boutros, E.~Nurvitadhi, and J.~Kim, ``A fast and flexible fpga-based accelerator for natural language processing neural networks,'' \emph{ACM Trans. Archit. Code Optim.}, vol.~20, no.~1, feb 2023. [Online]. Available: \url{https://doi.org/10.1145/3564606}
\BIBentrySTDinterwordspacing

\bibitem{hpta}
Y.~Han and Q.~Liu, ``Hpta: A high performance transformer accelerator based on fpga,'' in \emph{2023 33rd International Conference on Field-Programmable Logic and Applications (FPL)}, 2023, pp. 27--33.

\bibitem{swin}
Z.~Liu, P.~Yin, and Z.~Ren, ``An efficient fpga-based accelerator for swin transformer,'' 2023.

\bibitem{zhao_cao}
Z.~Zhao, R.~Cao, K.-F. Un, W.-H. Yu, P.-I. Mak, and R.~P. Martins, ``An fpga-based transformer accelerator using output block stationary dataflow for object recognition applications,'' \emph{IEEE Transactions on Circuits and Systems II: Express Briefs}, vol.~70, no.~1, pp. 281--285, 2023.

\bibitem{hw_fpga_ode}
I.~Okubo, K.~Sugiura, and H.~Matsutani, ``A cost-efficient fpga implementation of tiny transformer model using neural ode,'' 2024.

\bibitem{2024beta}
Y.~Ji, C.~Fang, and Z.~Wang, ``Beta: Binarized energy-efficient transformer accelerator at the edge,'' 2024.

\bibitem{marino2024mevit}
K.~Marino, P.~Zhang, and V.~Prasanna, ``Me-vit: A single-load memory-efficient fpga accelerator for vision transformers,'' 2024.

\bibitem{danopoulos2024transaxx}
D.~Danopoulos, G.~Zervakis, D.~Soudris, and J.~Henkel, ``Transaxx: Efficient transformers with approximate computing,'' 2024.

\bibitem{okubo2024costefficient}
I.~Okubo, K.~Sugiura, and H.~Matsutani, ``A cost-efficient fpga implementation of tiny transformer model using neural ode,'' 2024.

\bibitem{Zhuang_2024}
\BIBentryALTinterwordspacing
J.~Zhuang, Z.~Yang, S.~Ji, H.~Huang, A.~K. Jones, J.~Hu, Y.~Shi, and P.~Zhou, ``Ssr: Spatial sequential hybrid architecture for latency throughput tradeoff in transformer acceleration,'' in \emph{Proceedings of the 2024 ACM/SIGDA International Symposium on Field Programmable Gate Arrays}, ser. FPGA ’24.\hskip 1em plus 0.5em minus 0.4em\relax ACM, Apr. 2024. [Online]. Available: \url{http://dx.doi.org/10.1145/3626202.3637569}
\BIBentrySTDinterwordspacing

\bibitem{TurboTransformer}
\BIBentryALTinterwordspacing
J.~Fang, Y.~Yu, C.~Zhao, and J.~Zhou, ``Turbotransformers: an efficient gpu serving system for transformer models,'' in \emph{Proceedings of the 26th ACM SIGPLAN Symposium on Principles and Practice of Parallel Programming}, ser. PPoPP '21.\hskip 1em plus 0.5em minus 0.4em\relax New York, NY, USA: Association for Computing Machinery, 2021, p. 389–402. [Online]. Available: \url{https://doi.org/10.1145/3437801.3441578}
\BIBentrySTDinterwordspacing

\bibitem{jaewan_choi}
J.~Choi, H.~Li, B.~Kim, S.~Hwang, and J.~H. Ahn, ``Accelerating transformer networks through recomposing softmax layers,'' in \emph{2022 IEEE International Symposium on Workload Characterization (IISWC)}, 2022, pp. 92--103.

\bibitem{2022_softmax}
------, ``Accelerating transformer networks through recomposing softmax layers,'' in \emph{2022 IEEE International Symposium on Workload Characterization (IISWC)}, 2022, pp. 92--103.

\bibitem{2022_lightseq2}
X.~Wang, Y.~Wei, Y.~Xiong, G.~Huang, X.~Qian, Y.~Ding, M.~Wang, and L.~Li, ``Lightseq2: Accelerated training for transformer-based models on gpus,'' in \emph{SC22: International Conference for High Performance Computing, Networking, Storage and Analysis}, 2022, pp. 1--14.

\bibitem{he2023simplifying}
B.~He and T.~Hofmann, ``Simplifying transformer blocks,'' 2023.

\bibitem{yang2023inference}
N.~Yang, T.~Ge, L.~Wang, B.~Jiao, D.~Jiang, L.~Yang, R.~Majumder, and F.~Wei, ``Inference with reference: Lossless acceleration of large language models,'' 2023.

\bibitem{vLLMs}
\BIBentryALTinterwordspacing
W.~Kwon, Z.~Li, S.~Zhuang, Y.~Sheng, L.~Zheng, C.~H. Yu, J.~Gonzalez, H.~Zhang, and I.~Stoica, ``Efficient memory management for large language model serving with pagedattention,'' in \emph{Proceedings of the 29th Symposium on Operating Systems Principles}, ser. SOSP '23.\hskip 1em plus 0.5em minus 0.4em\relax New York, NY, USA: Association for Computing Machinery, 2023, p. 611–626. [Online]. Available: \url{https://doi.org/10.1145/3600006.3613165}
\BIBentrySTDinterwordspacing

\bibitem{zhao2024alisa}
Y.~Zhao, D.~Wu, and J.~Wang, ``Alisa: Accelerating large language model inference via sparsity-aware kv caching,'' 2024.

\bibitem{ham2020a3}
T.~J. Ham, S.~J. Jung, S.~Kim, Y.~H. Oh, Y.~Park, Y.~Song, J.-H. Park, S.~Lee, K.~Park, J.~W. Lee, and D.-K. Jeong, ``A$^3$: Accelerating attention mechanisms in neural networks with approximation,'' 2020.

\bibitem{2021_elsa}
T.~J. Ham, Y.~Lee, S.~H. Seo, S.~Kim, H.~Choi, S.~J. Jung, and J.~W. Lee, ``Elsa: Hardware-software co-design for efficient, lightweight self-attention mechanism in neural networks,'' in \emph{2021 ACM/IEEE 48th Annual International Symposium on Computer Architecture (ISCA)}, 2021, pp. 692--705.

\bibitem{2021_spatten}
H.~Wang, Z.~Zhang, and S.~Han, ``Spatten: Efficient sparse attention architecture with cascade token and head pruning,'' 2021.

\bibitem{2021_sanger}
\BIBentryALTinterwordspacing
L.~Lu, Y.~Jin, H.~Bi, Z.~Luo, P.~Li, T.~Wang, and Y.~Liang, ``Sanger: A co-design framework for enabling sparse attention using reconfigurable architecture,'' in \emph{MICRO-54: 54th Annual IEEE/ACM International Symposium on Microarchitecture}, ser. MICRO '21.\hskip 1em plus 0.5em minus 0.4em\relax New York, NY, USA: Association for Computing Machinery, 2021, p. 977–991. [Online]. Available: \url{https://doi.org/10.1145/3466752.3480125}
\BIBentrySTDinterwordspacing

\bibitem{salo}
\BIBentryALTinterwordspacing
G.~Shen, J.~Zhao, Q.~Chen, J.~Leng, C.~Li, and M.~Guo, ``Salo: an efficient spatial accelerator enabling hybrid sparse attention mechanisms for long sequences,'' in \emph{Proceedings of the 59th ACM IEEE Design Automation Conference}, ser. DAC '22.\hskip 1em plus 0.5em minus 0.4em\relax New York, NY, USA: Association for Computing Machinery, 2022, p. 571–576. [Online]. Available: \url{https://doi.org/10.1145/3489517.3530504}
\BIBentrySTDinterwordspacing

\bibitem{AccelTran}
S.~Tuli and N.~K. Jha, ``Acceltran: A sparsity-aware accelerator for dynamic inference with transformers,'' \emph{IEEE Transactions on Computer-Aided Design of Integrated Circuits and Systems}, vol.~42, no.~11, pp. 4038--4051, 2023.

\bibitem{DTQAtten1}
T.~Yang, D.~Li, Z.~Song, Y.~Zhao, F.~Liu, Z.~Wang, Z.~He, and L.~Jiang, ``Dtqatten: Leveraging dynamic token-based quantization for efficient attention architecture,'' in \emph{2022 Design, Automation and Test in Europe Conference and Exhibition (DATE)}, 2022, pp. 700--705.

\bibitem{2023_energon}
Z.~Zhou, J.~Liu, Z.~Gu, and G.~Sun, ``Energon: Toward efficient acceleration of transformers using dynamic sparse attention,'' \emph{IEEE Transactions on Computer-Aided Design of Integrated Circuits and Systems}, vol.~42, no.~1, pp. 136--149, 2023.

\bibitem{h3d}
\BIBentryALTinterwordspacing
Y.~Luo and S.~Yu, ``H3d-transformer: A heterogeneous 3d (h3d) computing platform for transformer model acceleration on edge devices,'' \emph{ACM Trans. Des. Autom. Electron. Syst.}, vol.~29, no.~3, apr 2024. [Online]. Available: \url{https://doi.org/10.1145/3649219}
\BIBentrySTDinterwordspacing

\bibitem{salo2}
J.~Zhao, P.~Zeng, G.~Shen, Q.~Chen, and M.~Guo, ``Hardware-software co-design enabling static and dynamic sparse attention mechanisms,'' \emph{IEEE Transactions on Computer-Aided Design of Integrated Circuits and Systems}, pp. 1--1, 2024.

\bibitem{2020_att}
\BIBentryALTinterwordspacing
H.~Guo, L.~Peng, J.~Zhang, Q.~Chen, and T.~D. LeCompte, ``Att: A fault-tolerant reram accelerator for attention-based neural networks,'' in \emph{2020 IEEE 38th International Conference on Computer Design (ICCD)}.\hskip 1em plus 0.5em minus 0.4em\relax Los Alamitos, CA, USA: IEEE Computer Society, oct 2020, pp. 213--221. [Online]. Available: \url{https://doi.ieeecomputersociety.org/10.1109/ICCD50377.2020.00047}
\BIBentrySTDinterwordspacing

\bibitem{hw_retransformer}
X.~Yang, B.~Yan, H.~Li, and Y.~Chen, ``Retransformer: Reram-based processing-in-memory architecture for transformer acceleration,'' in \emph{2020 IEEE/ACM International Conference On Computer Aided Design (ICCAD)}, 2020, pp. 1--9.

\bibitem{hw_inmemory}
A.~F. Laguna, A.~Kazemi, M.~Niemier, and X.~S. Hu, ``In-memory computing based accelerator for transformer networks for long sequences,'' in \emph{2021 Design, Automation and Test in Europe Conference and Exhibition (DATE)}, 2021, pp. 1839--1844.

\bibitem{transpim}
M.~Zhou, W.~Xu, J.~Kang, and T.~Rosing, ``Transpim: A memory-based acceleration via software-hardware co-design for transformer,'' in \emph{2022 IEEE International Symposium on High-Performance Computer Architecture (HPCA)}, 2022, pp. 1071--1085.

\bibitem{hw_xformer}
S.~Sridharan, J.~R. Stevens, K.~Roy, and A.~Raghunathan, ``X-former: In-memory acceleration of transformers,'' 2023.

\bibitem{trancim}
F.~Tu, Z.~Wu, Y.~Wang, L.~Liang, L.~Liu, Y.~Ding, L.~Liu, S.~Wei, Y.~Xie, and S.~Yin, ``Trancim: Full-digital bitline-transpose cim-based sparse transformer accelerator with pipeline/parallel reconfigurable modes,'' \emph{IEEE Journal of Solid-State Circuits}, vol.~58, no.~6, pp. 1798--1809, 2023.

\bibitem{h3datten}
W.~Li, M.~Manley, J.~Read, A.~Kaul, M.~S. Bakir, and S.~Yu, ``H3datten: Heterogeneous 3-d integrated hybrid analog and digital compute-in-memory accelerator for vision transformer self-attention,'' \emph{IEEE Transactions on Very Large Scale Integration (VLSI) Systems}, vol.~31, no.~10, pp. 1592--1602, 2023.

\bibitem{primate}
Y.~Pan, M.~Zhou, C.~Lee, Z.~Li, R.~Kushwah, V.~Narayanan, and T.~Rosing, ``Primate: Processing in memory acceleration for dynamic token-pruning transformers,'' in \emph{2024 29th Asia and South Pacific Design Automation Conference (ASP-DAC)}, 2024, pp. 557--563.

\bibitem{hardsea}
S.~Liu, C.~Mu, H.~Jiang, Y.~Wang, J.~Zhang, F.~Lin, K.~Zhou, Q.~Liu, and C.~Chen, ``Hardsea: Hybrid analog-reram clustering and digital-sram in-memory computing accelerator for dynamic sparse self-attention in transformer,'' \emph{IEEE Transactions on Very Large Scale Integration (VLSI) Systems}, vol.~32, no.~2, pp. 269--282, 2024.

\bibitem{key}
A.~Stillmaker and B.~Baas, ``Scaling equations for the accurate prediction of {CMOS} device performance from 180 nm to 7 nm,'' \emph{Integration, the {VLSI} Journal}, vol.~58, pp. 74--81, 2017, \url{http://vcl.ece.ucdavis.edu/pubs/2017.02.VLSIintegration.TechScale/}.

\bibitem{vhdl}
\BIBentryALTinterwordspacing
``Matrix multiplication in vhdl.'' [Online]. Available: \url{https://github.com/vangvassalos/MatMult\_VHDL}
\BIBentrySTDinterwordspacing

\end{thebibliography}


\end{document}